\documentclass{jpcdraft} %


\usepackage[round]{natbib}
\usepackage[ruled]{algorithm2e}

\usepackage[utf8]{inputenc} %
\usepackage[T1]{fontenc}    %
\usepackage{hyperref}       %
\usepackage{url}            %
\usepackage{booktabs}       %
\usepackage{amsfonts}       %
\usepackage{float}
\usepackage{amsmath}
\usepackage{amsthm}
\usepackage{amssymb}
\usepackage{mathrsfs}
\usepackage{graphicx}
\usepackage{nicefrac}       %
\usepackage{microtype}      %
\usepackage{xcolor}         %

\newtheorem{definition}{Definition}
\theoremstyle{definition}
\newtheorem{example}{Example}
\theoremstyle{plain} %

\newcommand{\sss}{\boldsymbol{s}}
\newcommand{\diffp}{\text{dp}}
\newcommand{\SSS}{\sss_\diffp}
\newcommand{\xx}{\boldsymbol{x}}
\newcommand{\uu}{\boldsymbol{u}}

\newcommand{\XX}{\mathcal{X}}
\newcommand{\RRR}{\mathbb{R}}
\def\ABC{ABC}

\begin{document}

\title[Exact Inference with Approximate Computation for DP Data]{Exact Inference with Approximate Computation for Differentially Private Data via Perturbations}

\author[R.~Gong]{Ruobin Gong}
\address{Department of Statistics, Rutgers University, New Brunswick, NJ 08854}
\email{rg915@stat.rutgers.edu}

\begin{abstract}
This paper discusses how two classes of approximate computation algorithms can be adapted, in a modular fashion, to achieve exact statistical inference from differentially private data products. Considered are approximate Bayesian computation for Bayesian inference, and Monte Carlo Expectation-Maximization for likelihood inference. Up to Monte Carlo error, inference from these algorithms is exact with respect to the joint specification of both the analyst's original data model, and the curator's differential privacy mechanism. Highlighted is a duality between approximate computation on exact data, and exact computation on approximate data, which can be leveraged by a well-designed computational procedure for statistical inference.
\end{abstract}

\maketitle

\begin{center}
	First version: September 2019 \qquad This version: September 2022
\end{center}

\section{Introduction}\label{sec:intro}

Differential privacy \citep{dwork2006calibrating} advances statistical disclosure limitation by putting forth a formal and practical framework. In addition to grounding the concept of privacy on a mathematical footing, differential privacy distinguishes itself from traditional approaches by offering transparent probabilistic mechanisms, whose specifications can be made public without sabotaging the privacy guarantee. Differential privacy has been adapted by major data curators in the industry, research and government sectors. 
As a prime example, the U.S. Census Bureau deploys differential privacy to protect the 2020 Decennial Census data products \citep{abowd2022topdown}. The P.L. 94-171 redistricting data files have been released on August 12, 2021 \citep{census2021PL}.

In this work, we adopt the perspective of a data analyst operating under the \emph{dissemination} mode of data access \citep{hotz2022balancing}. A data curator, such as the Census Bureau, collects potentially sensitive data and releases differentially private data products to the analyst. The analyst in turn conducts statistical inference for their quantities of interest based on the privatized data. The analyst's goal is to draw trustworthy inference from the statistical model they wish to fit, knowing that the data have undergone privacy protection. This may not be a trivial task. The curator instills differential privacy in the data product via a data processing mechanism. Na\"ively treating processed data as if they are unprocessed may result in erroneous and misleading statistical inference. With the wide adoption of differential privacy for disclosure limitation, social scientists and policy researchers are faced with the challenge to revise their preferred statistical analyses to account for the privacy mechanism, however complex they may be. To keep up with advances in privacy protection, we  need theoretically sound and computationally efficient statistical methodologies to supplant their predecessors \citep{hansen_2018}.

This paper discusses the adaptation of two classes of approximate computation algorithms, \emph{approximate Bayesian computation}  (\ABC) and \emph{Monte Carlo Expectation-Maximization} (MCEM), to obtain exact Bayesian and likelihood statistical inferences based on differentially private data products. The word \emph{exact} means that, up to Monte Carlo error, the resulting inference corresponds precisely to the joint statistical model that accounts for both the analyst's specifications and the differential privacy mechanism. This paper draws a concrete connection between the novel disclosure limitation mechanisms that obey differential privacy, and the vast reserve of computational strategies available for likelihood and Bayesian statistical inference. The hope is that users of traditional, non-differentially private data can smoothly transition their existing methodologies to suit novel, differentially private data products while maintaining statistical validity. The two methods discussed in this work are applicable to a wide range of existing models, dispensing the need to analytically recompute the new joint model to account for the privacy mechanism.  Both classes of algorithms discussed in this paper do not assume specific structures of the likelihood, prior, and privacy mechanism. Indeed, the likelihood approach only requires that the analyst's original model is suitable for EM, and the Bayesian approach only requires that the original likelihood can be simulated and that the prior is proper. Should specific and convenient model structures be available, the proposed mechanisms would be amenable to adaptation to enhance computational efficiency.

The remainder of this paper is organized as follows. Section~\ref{subsec:dp} lays out the mathematical formalism and notation for differential privacy and perturbation mechanisms. Section~\ref{sec:dpabc} proposes a rejection \ABC~algorithm, and shows that with kernel and bandwidth chosen to correspond to the perturbation mechanism underlying the privatized data, it produces exact posterior inference in the form of independent and identically distributed samples from the true posterior distribution. Section~\ref{sec:mcem} discusses an importance sampling implementation of Monte Carlo EM for likelihood inference. The validity of both approximate computation methods owes to the fact that their tuning elements can be chosen in accordance with the differentially private perturbation mechanism that is used to generate the privatized data product. Section~\ref{sec:example} provides two numerical demonstrations, on the Bayesian and likelihood inference for privatized count, and a differentially private adaptation of the Lalonde dataset for inference on job training program efficacy. Section~\ref{sec:discussion} concludes with a discussion on the duality between \emph{approximate computation on exact data} and \emph{exact computation on approximate data}, and the various challenges to the efficiency of these proposals.

\section{Differential privacy and perturbation mechanism}\label{subsec:dp}

Differential privacy aims to protect the confidential information of individual respondents in a dataset $\xx \in \XX$, without undue sacrifice of accuracy in learning about aggregate features of the underlying population as represented by $\xx$. Here, an aggregate feature is a query $\sss: \XX \to \RRR^p$, a deterministic function of $\xx$, such as the sample average, variance, quantiles and so on. Queries are the means through which analysts learn from the dataset. Counting queries, including histograms and contingency tables which are ordered multivariate counts over a partition of $\xx$, constitute a most useful class of queries. It is the main query type for the 2020 U.S. Census data products, tabulated across various geographic levels such as states, counties, and Census blocks.

Differential privacy is realized via a probabilistic mechanism based on the intended query. A differentially private query reflects as truthful as possible about the status of $\xx$, meanwhile behaves similarly should it be calculated based on any neighboring dataset of $\xx$. The notion of differentially privacy is defined in probabilistic terms.

\begin{definition}[differential privacy; \citealp{dwork2006calibrating}]\label{def:dp2}
A random function $\SSS:\XX\to\RRR^{p}$ is $\left(\epsilon,\delta\right)$-differentially private if for all neighboring datasets $ \left(\xx,\xx'\right) \in \XX^{2}$ and all $A \in \mathscr{B}(\RRR^p)$, 
\begin{equation}\label{eq:ed-dp}
	Pr\left(\SSS\left(\xx'\right) \in A\right) \le e^{\epsilon} \cdot Pr\left(\SSS\left(\xx\right) \in A\right) + \delta.
\end{equation} 
$\SSS$ is $\epsilon$-differentially private if it is $\left(\epsilon,0\right)$-differentially private.
\end{definition}

The pair $\left(\xx, \xx'\right) \in \XX^{2}$ constitutes neighboring datasets if they differ by precisely one entry, either by adding or dropping one respondent, or by taking a different value (as used in the definition of \emph{bounded} differential privacy; \citealp{dwork2006calibrating}). When operating on neighboring datasets, the random function $\SSS$ induces pairs of probability measures, associated respectively with  $\SSS(\xx)$ and $\SSS(\xx')$, that are close to each other. The degree of closeness is controlled by the \emph{privacy loss budget} $\epsilon$ and $\delta$. In the extreme case that both are zero, the two measures must be equal on every Borel set $A$, which for general $\xx$ can only happen if $\SSS$ does not depend on the data at all. In other words, differential privacy requires that the distribution of $\SSS$ to be stable within the small neighborhood around the observable dataset.

Differential privacy is a property pertinent to the random function $\SSS$. Many widely employed differential privacy mechanisms take the form of perturbation mechanisms.

\begin{definition}\label{def:pert}
For a dataset $\xx \in \XX$ and a deterministic function $\sss: \XX \to \RRR^{p}$, the random function $\SSS$ is a \emph{perturbation mechanism} based on $\sss$ if
\begin{equation}\label{eq:dp-mechanism}
\SSS\left(\xx\right) \mid \sss\left(\xx\right) \sim \eta_{\diffp}\left(\;\cdot\; \mid \sss\left(\xx\right) \right),
\end{equation}
for $\eta_{\diffp}$ a known conditional probability distribution. In particular, $\SSS$ is an \emph{additive perturbation mechanism} based on $\sss$ if
\begin{equation}\label{eq:pert}
	\SSS\left(\xx\right) = \sss\left(\xx\right) + h\uu,
\end{equation}
where the noise component $\uu$ is a $p$-dimensional random variable with known distribution $\eta$, $\mathbb{E}(\uu) = {\bf 0}$, and $h > 0$ is a scale  (or bandwidth) parameter.	
\end{definition}

The differentially private query $\SSS$ is a noisy version of its deterministic counterpart $\sss$. The protection of privacy is achieved through randomly perturbing what would otherwise be a deterministic query calculated based on $\xx$. The subscript ``\diffp'' in $\SSS$ emphasizes that it instantiates the privacy  mechanism $\eta_\diffp$, rather than the data generation mechanism of $\xx$, as the analyst might posit. The perturbation mechanism embodied by $\SSS$ is said to be \emph{unbiased} if it satisfies $\mathbb{E}\left(\SSS\left(\xx\right) \mid\sss\left(\xx\right) \right)=\sss\left(\xx\right)$. Additive perturbation mechanisms, by construction of~\eqref{eq:pert}, are unbiased. Furthermore, if the scale parameter $h$ does not depend on the confidential dataset $\xx$, the mechanism may be called a \emph{data-independent} mechanism \citep{li2015matrix}. Note that the additive perturbation mechanism resembles the classical measurement error model \citep{carroll2006measurement}, where the noise $h\uu$ has a known distribution, and the noisy measurement $\SSS$ is observed precisely once.  Appendix~\ref{app:dp-mechanism} gives three examples of widely used differentially private mechanisms, with additive perturbation using Gaussian and Laplace noises. Their definitions invoke three notions of \emph{functional sensitivity}, \eqref{eq:global}-\eqref{eq:local}, which we generally denote as $\Delta(\sss)$, to capture the idea that certain $\sss$ is more revealing of individual information in $\xx$ than others. It is crucial that the scale parameter of the additive perturbation mechanism is chosen as a function of both the sensitivity of $\sss$ and the privacy budget, i.e. $h = h(\epsilon,\delta,\Delta(\sss))$. Additional examples of additive differentially private mechanisms include the generalized Cauchy \citep{nissim2007smooth}, double Geometric \citep{schein2019locally}, correlated multivariate Gaussian \citep{nikolov2013geometry} and the $k$-norm \citep{hardt2010geometry,bhaskara2012unconditional} mechanisms. Examples of non-additive perturbation mechanisms include the randomized response mechanism \citep{warner1965randomized}, exponential mechanism \citep{mcsherry2007mechanism}, objective perturbation \citep{Chaudhuri2011,Kifer2012:PrivateCERM}, among others.

A primary strength of differential privacy over traditional disclosure limitation frameworks is its \emph{transparency}, which means that the specification of the perturbation mechanism $\eta_\diffp$ may be fully revealed to the data analyst (and indeed the public) while keeping the privacy guarantee intact. For additive mechanisms, this specification consists of $\uu$'s distribution $\eta$, scale parameter $h$, and the privacy loss budget $\epsilon$ and $\delta$. Perturbation mechanisms can be correctly accounted for in the probabilistic modeling of privatized data. Despite the necessary sacrifice of statistical efficiency, likelihood and Bayesian models utilizing privatized data can still retain validity, in the sense that any inference drawn based on $\sss$ can still be drawn based on $\sss_\diffp$ correctly while accounting for its generative process. As Section~\ref{sec:dpabc} will discuss, for Bayesian analysis, an \ABC~rejection algorithm guarantees the exactness of draws from the true posterior distribution, when properly tuned according to the parameters of the perturbation mechanism. The  nature of the privatized query makes \ABC~an appealing choice for posterior computation, even when the model is not as complex as to necessitate its use.

\section{Exact Bayesian inference with differentially private data}\label{sec:dpabc}

In the absence of privacy protection, suppose a Bayesian model was posited based on the confidential query $\sss$ as a function of $\xx$. Let  $\sss\left(\xx\right)\mid\theta\sim \pi(\sss \mid \theta)$ be the confidential data likelihood, and $\theta\sim\pi_{0}(\theta)$ the prior distribution for $\theta$. The posterior distribution of $\theta$ given $\sss$ is 
\begin{equation}\label{eq:noiseless-posterior}
\pi\left(\theta \mid \sss\right) \propto \pi_{0}\left(\theta\right)\pi\left(\sss\mid\theta\right).
\end{equation}
If the query $\sss$ isn't privacy-protected, quantities calculated based on~\eqref{eq:noiseless-posterior}, either analytically or via simulation, would conclude the Bayesian analysis. With the privacy protection mechanism in place, however, we no longer observe the confidential query $\sss$, but rather the  privatized (perturbed) query $\sss_\diffp$ as a single realization of the privacy mechanism~\eqref{eq:dp-mechanism}. The joint distribution of $\theta$ and $\sss_\diffp$ is 
\[
\pi\left(\theta,\sss_{\diffp}\right)=\int\pi\left(\theta,\sss\right)\eta_\diffp\left(\sss_{\diffp}\mid\sss\right)d\sss,
\]
marginalized over the latent $\sss$. This identity holds because the conditional distribution of $\sss_\diffp$ given $\sss$ and $\theta$ is free of $\theta$, as it is precisely the known perturbation mechanism: $\pi\left(\sss_{\diffp}\mid\sss, \theta \right) = \eta_\diffp\left(\sss_{\diffp}\mid\sss\right)$. The posterior distribution of $\theta$ given $\sss_\diffp$ is
\begin{equation}\label{eq:posterior}
\pi\left(\theta\mid\sss_{\diffp}\right) = \int\frac{\pi\left(\sss,\sss_{\diffp},\theta\right)}{\pi\left(\sss_{\diffp}\right)}d\sss = \frac{\pi_0\left(\theta\right)\int\eta_\diffp\left(\sss_{\diffp} \mid \sss\right)\pi\left(\sss\mid\theta\right)d\sss}{\int\pi_0\left(\theta\right)\int\eta_\diffp\left(\sss_{\diffp} \mid \sss\right)\pi\left(\sss\mid\theta\right)d\sss d\theta}.
\end{equation}
As~\eqref{eq:posterior} is the true posterior distribution for $\theta$ given the observable information,  analytical or simulated computation based on~\eqref{eq:posterior} would conclude the exact Bayesian analysis. However, computation of~\eqref{eq:posterior} may not be trivial, as part of it involves the observed likelihood $\int\eta_\diffp\left(\sss_{\diffp} \mid \sss\right)\pi\left(\sss\mid\theta\right)d\sss$, which is an integral of the product between the confidential data likelihood and the privacy mechanism. The challenge is exacerbated by the fact that the confidential likelihood is specified by the data analyst, whereas the privacy mechanism is specified by the data curator. These choices are typically made independent of one another, and either of them may already be complex and computationally demanding on its own. 

Algorithm~\ref{algorithm:abc} presents a recipe to generate independent and identically distributed samples from the exact posterior distribution~\eqref{eq:posterior}. It demands little of the tractability of the confidential likelihood. The only requirement is that for given values of $\theta$, one can simulate from $\pi(\sss\mid\theta)$, but otherwise it need not be available in closed form. Algorithm~\ref{algorithm:abc} is a type of \ABC~algorithm, which was designed to supply practical solutions to large-scale models for which the likelihood may be implicit or intractable and posteriors without closed-form expressions. \ABC~brought computational feasibility to stochastic differential equation models for complex dynamic systems in population genetics \citep{beaumont2002approximate}, systems biology \citep{toni2008approximate} and ecology \citep{wood2010statistical}, albeit~\ABC~posteriors are typically only approximate relative to the true target posterior. However, as will be shown in Theorem~\ref{thm:dp-abc} and discussed in Section~\ref{sec:discussion}, the employment of \ABC~for differentially private data serendipitously eradicates the ``approximate'' nature of the resulting posterior samples, which otherwise would be the case if the data were noise-free.

\begin{algorithm}[t]
 \KwData{Privatized query $\sss_\diffp$, perturbation mechanism  $\eta_{\diffp}$;}
 \KwResult{A set of parameter values $\{\theta_i\}_{i=1}^{N}$;}
 \For{each $i=1,\ldots,N$}{
  1. Simulate $\theta_i \sim \pi_0(\theta)$\;
  2. Simulate $\sss_i \sim \pi(\sss \mid \theta_i)$\;
  3. Accept $\theta_i$ with probability $c\eta_{\diffp}\left(\sss_{\diffp} \mid \sss_i\right)$ where $c^{-1}=\max \eta_{\diffp}$, otherwise go to step 1\;
 }
 \caption{Rejection \ABC~algorithm with differentially private queries}
 \label{algorithm:abc}
\end{algorithm}

\begin{thm}\label{thm:dp-abc} 
Let $\pi(\sss\mid\theta)$ be the likelihood for the unobserved confidential query $\sss$, $\pi_{0}(\theta)$  a proper prior distribution, and  $\eta_{\diffp}\left(\sss_\diffp  \mid \sss\right)$ a perturbation mechanism. Then, Algorithm~\ref{algorithm:abc} samples independently and identically from the exact posterior distribution $\pi\left(\theta \mid \sss_{\diffp}\right)$ defined in \eqref{eq:posterior}.
\end{thm}

Proof of Theorem~\ref{thm:dp-abc} can be found in Appendix~\ref{app:dp-abc}. Key to the validity of  Theorem~\ref{thm:dp-abc} is that the differentially private perturbation mechanism is ignorable for $\theta$ \citep{little2014statistical}, or in other words, the unobserved confidential query $\sss$ is sufficient with respect to the complete likelihood $\pi(\sss,\sss_\diffp\mid \theta)$.  
Traditional statistical disclosure limitation mechanisms may or may not enjoy ignorability, a matter further complicated by their non-transparency to impact the quality of downstream statistical analysis  \citep{abowd2016economic}. By contrast, the ignorability property of differential privacy enables exact statistical inference and may substantially simplify the computational task.

An intuitive connection with traditional \ABC~can be drawn if we restrict attention to the case of additive perturbation. As defined in~\eqref{eq:pert}, assume $\eta_\diffp\left(\sss_{\diffp}\mid\sss\right) = \eta\left(\left(\sss_{\diffp}-\sss\right)/h\right)$ where $\eta(\cdot)$ is the density of the additive noise $\uu$ and $h$ a scale parameter, both known precisely to the analyst. Algorithm~\ref{algorithm:abc} adopts the kernel density $\eta$, properly scaled by a factor of $c$, with bandwidth $h = h(\epsilon,\delta,\Delta(\sss))$ and center $\sss_\diffp$ to be its acceptance probability at step 3, thus reduces to a classic rejection \ABC~algorithm with tuning parameters (i.e. kernel and bandwidth) set to match precisely the additive perturbation mechanism employed to generate $\sss_\diffp$.

One way to understand Theorem~\ref{thm:dp-abc} is that the privacy mechanism plays the role of the ``random summary statistic'' in the noisy \ABC~algorithm of \cite{fearnhead2012constructing}. Noisy \ABC~is calibrated with respect to the joint Bayesian model, whereas \ABC~typically isn't. However, the kernel and bandwidth in noisy \ABC~are merely parameters to fine-tune the tradeoff between approximation error and the Monte Carlo error in the posterior, which in turn controls the efficiency of the sampler. In contrast, both the kernel and the bandwidth of Algorithm~\ref{algorithm:abc} are dictated externally by the perturbation mechanism and the privacy loss budget. The computational tradeoff and the privacy tradeoff are ``bundled'' together: specifying the parameters of \ABC~also specifies those of the privacy mechanism, and vice versa.

The overall acceptance probability of Algorithm~\ref{algorithm:abc} is $\pi\left(\sss_{\diffp}\right)/\max{\eta_\diffp}$, or the model evidence evaluated at $\sss_\diffp$ divided by the modal density of $\eta_\diffp$  (see Appendix~\ref{app:dp-abc}). This means that rejection can be frequent if model evidence is low, such as when the prior and the observed likelihood are in disagreement (termed \emph{prior-data conflict}; \citealp{evans2006checking}), or if the privacy bandwidth $h$ is too small.

To address the concern, Algorithm~\ref{algorithm:abc} can be adapted to work with a variety of alternative \ABC~sampling techniques to produce consistent posterior estimates for functions of interest. As an example, we discuss an importance sampling variation to Algorithm~\ref{algorithm:abc} as follows. At step 1 of each iteration, sample $\theta_i \sim g(\theta)$, a proposal distribution that is positive wherever the prior $\pi_0(\theta)$ is positive. At step 3, no rejection is performed, but instead $\theta_i$ is assigned a weight 
\[
\omega_i = \omega(\sss_i,\theta_i) = \eta_\diffp\left(\sss_{\diffp} \mid \sss_i\right)\pi_0(\theta_i)/g(\theta_i).
\]
The algorithm returns weighted draws $\{\theta_i, \omega_i\}_{i=1}^{N}$. For a square-integrable function of interest $a(\theta)$, the weighted average estimator converges in probability to its posterior expectation given $\sss_\diffp$ as $N \to \infty$ \citep{liu2008monte}:
\begin{equation}\label{eq:consistency}
\frac{\sum_{i=1}^{N}\omega_{i}a\left(\theta_{i}\right)}{\sum_{i=1}^{N}\omega_{i}}\overset{p}{\to}\frac{\mathbb{E}_{g}\left(\omega\left(\theta,\sss\right)a\left(\theta\right)\right)}{\mathbb{E}_{g}\left(\omega\left(\theta,\sss\right)\right)}=\mathbb{E}\left(a\left(\theta\right)\mid\sss_{\diffp}\right),	
\end{equation} 
where $\mathbb{E}_{g}\left(\cdot\right)$ is with respect to the joint distribution $g(\theta)\pi(\sss\mid\theta)$, and $\mathbb{E}\left(\cdot\mid\sss_{\diffp}\right)$ is with respect to the true posterior in (\ref{eq:posterior}). The proposal distribution $g(\cdot)$ can be chosen to minimize the variance of the estimator in \eqref{eq:consistency}, such as a density that is close in shape to $a(\theta)\pi_0(\theta)$ \citep{liu2008monte}. Further adaptations of and beyond~\ABC, such as hybrid importance-rejection sampling \citep{fearnhead2012constructing}, rejection control \cite[ch.4]{sisson2018handbook}, Markov chain Monte Carlo \citep{marjoram2003markov} and sequential Monte Carlo \citep{sisson2007sequential} can be developed likewise, while the consistency result of \eqref{eq:consistency} remains standing. %

\section{Exact likelihood inference with differentially private data}\label{sec:mcem}

This section discusses a Monte Carlo Expectation-Maximization \citep[EM;][]{dempster1977maximum, wei1990monte} implementation for likelihood inference with differentially private data. Under the classic setting, when a likelihood involves both observed and latent data, EM seeks the maximum likelihood estimate of the parameter by iteratively integrating the log likelihood over the conditional predictive distribution of the latent data given the observed data and a current parameter value (the E-step), and maximizing the parameter value over this integral (the M-step).

In the context of differential privacy, the complete data is $(\sss, \sss_\diffp)$, in which the latent data is the confidential query $\sss$, and the observed data is the privatized query $\sss_\diffp$. In the special case of additive perturbation, $\sss_\diffp = \sss + h\uu$ is a convolution of $\sss$ and the noise component $\uu$. The complete likelihood is written as $L(\theta;\sss,\sss_{\diffp}) \propto \pi(\sss,\sss_\diffp\mid \theta)$, as defined in Section~\ref{sec:dpabc}. The EM algorithm for maximum likelihood inference for $\theta$ given the differentially private $\sss_\diffp$ is schematically described in Algorithm~\ref{algorithm:em}. 

\begin{algorithm}[H]
\KwData{Privatized query $\sss_\diffp$, initial $\theta^{\left(0\right)}$;}
\KwResult{A local maximizer $\theta^{\left(t^{*}\right)}$;}
\While{$\Delta\left(\theta^{\left(t\right)},\theta^{\left(t-1\right)}\right) > $tol.}{
\emph{E-step}: Evaluate the expectation of the complete log likelihood with respect to the conditional predictive distribution of $\sss$ given $\sss_\diffp$ and the current maximizer $\theta^{(t)}$:
\begin{eqnarray}
Q(\theta; \theta^{(t)}) &=& \mathbb{E}\left(
 \log L(\theta;\sss,\sss_{\diffp})\mid\sss_{\diffp},\theta^{(t)}\right) \nonumber\\
 &=&\mathbb{E}\left( \log\pi(\sss\mid\theta)\mid\sss_{\diffp},\theta^{(t)}\right) +\text{const.}; \label{eq:q-function}
\end{eqnarray}

  \emph{M-step:}  Calculate $\theta^{(t+1)} := \text{argmax}_{\theta} Q(\theta;  \theta^{(t)})$, and set $t := t+1$\;
 }
 \caption{EM algorithm for differentially private queries}
\label{algorithm:em}
\end{algorithm}
The E- and M-steps are iterated until convergence, that is when $\theta^{(t)}$ stabilizes so that its distance (somehow measured) from the previous iteration, $\text{dist}\left(\theta^{\left(t\right)},\theta^{\left(t-1\right)}\right)$, is sufficiently small. It is worth noting that the constant term in \eqref{eq:q-function} is equal to 
\[
\mathbb{E}\left( \log \eta_\diffp(\sss_{\diffp}\mid\sss)\mid\sss_{\diffp},\theta^{(t)}\right),
\]
which can be ignored within the EM algorithm. This is because, as discussed in Section~\ref{sec:dpabc}, the privacy mechanism $\eta_\diffp$ is known and independent of $\theta$, and so is the conditional predictive expectation of its log density.

As alluded to in Section~\ref{sec:intro}, for likelihood modeling of differentially private data, the confidential data likelihood and the privacy mechanism are typically specified by separate parties without coordination with one another. Thus in general, one cannot expect the observed data likelihood (which is an integral of their product) to come from an exponential family \citep[cf.][]{park2017dp}, nor be able to perform both the E- and the M-steps analytically. Monte Carlo implementation of one or both steps may be needed, which amounts to implementing the E-step of Algorithm~\ref{algorithm:em} via an importance sampling scheme. We  describe this scheme in Algorithm~\ref{algorithm:mcem}. The set of weighted samples $\{\sss_i, \omega_i\}_{i=1}^{N}$ produced by Algorithm~\ref{algorithm:mcem} may be used in two ways, depending on whether the confidential data likelihood is or isn't an exponential family. We discuss both cases below.

\begin{algorithm}[H]
 \KwData{Privatized query $\sss_\diffp$, perturbation mechanism $\eta_{\diffp}$;}
 \KwResult{A set of weighted samples $\{\sss_i, \omega_i\}_{i=1}^{N}$, to be used for~\eqref{eq:ets}-\eqref{eq:mcem-fisher};}
 \For{the $t^{th}$ E-step of Algorithm~\ref{algorithm:em},}{
  1. Simulate $\sss_i \sim \pi(\sss \mid \theta^{(t)})$\;
  2. Calculate $\omega_i = \eta_\diffp\left( \sss_{\diffp} \mid \sss_i \right)$\;
 }
 \caption{E-step via importance sampling for differentially private queries}
\label{algorithm:mcem}
\end{algorithm}

\subsection{Confidential data with exponential family likelihood.}

In the simpler scenario that the confidential data likelihood $\pi(\sss \mid \theta)$ as specified by the analyst belongs to the exponential family, it admits a sufficient statistic to the parameter $\theta$ which we denote as $b\left(\sss\right)$.  The function $Q(\theta; \theta^{(t)})$ in  (\ref{eq:q-function}) can be written as an explicit function of $\theta$ and 
\begin{equation}\label{eq:ets}
\mathbb{E}\left( b(\sss)\mid\sss_{\diffp},\theta^{(t)}\right),
\end{equation}	
the conditional expectation of $b(\sss)$ given $\sss_{\diffp}$ and the current maximizer $\theta^{(t)}$. With this simplification, however, (\ref{eq:ets}) may still not be evaluable in closed form, in which case we utilize the set of weighted samples $\{\sss_i, \omega_i\}_{i=1}^{N}$ produced by~Algorithm~\ref{algorithm:mcem} to consistently estimate it at every iteration $t$. Indeed, as $N \to \infty$, the weighted estimator
\begin{equation}\label{eq:ets-estimator}
\frac{\sum_{i=1}^{N}\omega_{i}b\left(\sss_{i}\right)}{\sum_{i=1}^{N}\omega_{i}}
\end{equation}
converges in probability to (\ref{eq:ets}). For the E-step of the $(t+1)$st iteration, $\theta^{(t+1)}$ can be found by maximizing $Q(\theta; \theta^{(t)})$, replacing (\ref{eq:ets}) therein with \eqref{eq:ets-estimator}. The
 effective sample size at the $t$th iteration is
\begin{equation}\label{eq:mcem-ess}
\textsc{ess}^{\left(t\right)}\left(N\right) =  N\pi^{2}\left(\sss_{\diffp}\mid\theta^{(t)}\right)\mathbb{E}^{-1}_{\sss\mid\theta^{\left(t\right)}}\left( \eta^{2}_\diffp\left( \sss_{\diffp} \mid \sss \right) \right),
\end{equation} 
where the subscript ``$\sss\mid\theta^{\left(t\right)}$'' signifies that the expectation is taken with respect to the current approximation to the confidential data likelihood, or equivalently, the proposal distribution of the E-step importance sampler. Derivation of \eqref{eq:mcem-ess} may be found in Appendix~\ref{app:mcem-weights}.

In Algorithm~\ref{algorithm:mcem}, the $\sss_i$'s are simulated from the current approximation to the analyst's confidential data likelihood, and the weights $\omega_i$'s are separately determined by the curator's privacy mechanism. Similar in spirit to Algorithm~\ref{algorithm:abc}, this separation allows the computation to easily accommodate independently derived choices of data likelihood and privacy mechanisms, and does not require the evaluation or integration of quantities that are nontrivial functions of both.
Whenever appropriate, however, one may modify Algorithm~\ref{algorithm:mcem} to sample from the conditional predictive distribution in more efficient ways. For example, with rejection or Markov chain-based samplers, $\sss_i$ follows a proposal distribution and $\omega_i = 1$ if $\sss_i$ is accepted and $0$ otherwise \citep{mcculloch1997maximum, booth1999maximizing}. One may also perform importance sampling where $\sss_i \sim \pi\left( \sss \mid\sss_{\diffp},\theta^{(t-1)}\right)$, the approximation to the conditional predictive distribution at the previous iteration, and $\omega_i = \pi\left( \sss \mid\sss_{\diffp},\theta^{(t)}\right)/\pi\left( \sss \mid\sss_{\diffp},\theta^{(t-1)}\right)$ the ratio between the current and previous approximations, thereby reweighting and recycling the multiply-imputed $\sss_i$'s to save computational effort \citep{quintana1999monte}. One may also resample the simulated $\sss_i$'s according to their associated weights to obtain an unweighted rejection sample, as long as the goal is to construct as accurate as possible an estimate for \eqref{eq:ets} as part of the E-step.

\subsection{Confidential data with general likelihood.}
If the confidential data likelihood does not come from an exponential family, $Q(\theta; \theta^{(t)})$ of (\ref{eq:q-function}) may not reduce to a straightforward expression involving $\theta$ and (\ref{eq:ets}). In this case, the E-step requires a full approximation to $Q(\theta; \theta^{(t)})$ as a mixture of augmented log likelihoods, constructed as follows. 

Let $\{\sss_i,\omega_i\}_{i=1}^N$ be a weighted sample from the target distribution $\pi\left( \sss \mid\sss_{\diffp},\theta^{(t)}\right)$, the $t$th approximation to the conditional predictive distribution. Specifically $\{\sss_i,\omega_i\}_{i=1}^N$ can be the importance sample generated by Algorithm~\ref{algorithm:mcem}, or by one of its variations such as as described above. Then,
\begin{equation}\label{eq:mcem-q}
\hat{Q}(\theta ;  \theta^{(t)}) = m{\sum_{i=1}^{N}\omega_{i}\log\pi(\sss_{i}\mid\theta)}
\end{equation}
serves as a consistent approximation to $Q(\theta; \theta^{(t)})$. The constant $m^{-1} = \sum_{i=1}^{N}\omega_{i}$ in (\ref{eq:mcem-q}) is inconsequential to the maximizer in the ensuing M-step, as long as the $\omega_i$'s do not involve the unknown parameter $\theta$. That is indeed the case since, again, the perturbation mechanism is ignorable for $\theta$. Writing $\lambda_\theta(\sss) = \nabla_\theta\log\pi\left(\sss \mid\theta\right)$, the observed score function  $\nabla_\theta \log\pi\left(\sss_{\diffp}\mid\theta^{(t)} \right)$  can be approximated at the $t$th iteration according to 
\begin{equation}\label{eq:mcem-score}
\mathbb{E}\left( \lambda_\theta(\sss) \mid\sss_{\diffp},\theta^{\left(t\right)}\right)\overset{\cdot}{=}  m\sum_{i=1}^{N}\omega_{i}\lambda_\theta(\sss_i).
\end{equation}
The observed Fisher information can also be approximated according to 
\begin{multline}\label{eq:mcem-fisher}
-\nabla_\theta^{2}\log\pi\left(\sss_{\diffp}\mid\theta^{(t)}\right)  \overset{\cdot}{=} \\
m\sum_{i=1}^{N}\omega_{i}\left\{ -\nabla_{\theta}\lambda_{\theta}\left(\sss_{i}\right)-\lambda_{\theta}\left(\sss_{i}\right)\lambda_{\theta}\left(\sss_{i}\right)^{\top}\right\} + m^{2}\sum_{i=1}^{N}\sum_{j=1}^{N}\omega_{i}\omega_{j}\lambda_{\theta}\left(\sss_{i}\right)\lambda_{\theta}\left(\sss_{j}\right)^{\top}. 	
\end{multline}
Derivations of the observed score function and observed Fisher information can be found in Appendix~\ref{app:mcem-details}. Both~\eqref{eq:mcem-score} and~\eqref{eq:mcem-fisher} may be used for quantifying the inferential uncertainty under the normal approximation to the likelihood \citep{meilijson1989fast}, as well as accelerating and assessing convergence for Newton-type implementations of the M-step. The approximations given above rely only on that the first and second derivatives of the confidential likelihood be evaluable at the simulated $s_i$'s.

For any EM algorithm (and not just Monte Carlo EM) to be applicable to likelihood inference from differentially private data, one must be able to evaluate the confidential data likelihood $\pi(\sss \mid \theta)$, to the extent that maximization of the $Q$ function can be done at least numerically. 
The vast literature on Monte Carlo EM has much to instruct on implementing both the E- and the M-steps with better convergence rates, sampling efficiency, or under computational capacity constraints, for adapting modeling scenarios to differentially private data. The additive perturbation mechanism of \eqref{eq:pert} is a special instance of a linear mixed effects model, which is particularly suitable for Monte Carlo EM and has been studied extensively in the literature, e.g. \cite{wolfinger1993generalized,mcculloch1997maximum}.

\section{Numerical Demonstrations}\label{sec:example}

\subsection{Bayesian and likelihood inference from privatized count}\label{sec:example-count}

In this simple example, we consider modeling the number of respondents from a sample $\xx$ in possession of a certain feature. $\sss(\cdot)$ is the univariate counting query, for which we posit the sampling model $\sss\left(\xx\right)\mid\theta\sim Pois\left(\theta\right)$. $\theta$ is the population expectation parameter for which we wish to draw Bayesian and likelihood inference. 

First consider a Bayesian model for $\theta$. We implement rejection~\ABC~as described in Algorithm~\ref{algorithm:abc} to draw from the exact Bayesian posterior based on the privatized count $\sss_\diffp$. Suppose $\sss_\diffp$  is produced by the $\epsilon$-differentially private Laplace mechanism  (Example~\ref{ex:1} in Appendix~\ref{app:dp-mechanism}), where the additive noise follows $\uu \sim Lap_p(1)$ with bandwidth $h = \epsilon^{-1}$. As with general \ABC~samplers, Algorithm~\ref{algorithm:abc} can work with arbitrary choices of prior and likelihood that need not be conjugate, so long as the prior is proper. For the purpose of illustration, we consider the prior $\theta \sim Gamma\left(\alpha,\beta\right)$, where $\alpha$ and $\beta$ are fixed hyperparameters, so that an analytically tractable posterior can be obtained for visual comparison.

\begin{figure}[t]
\centering
\includegraphics[width = .6\textwidth]{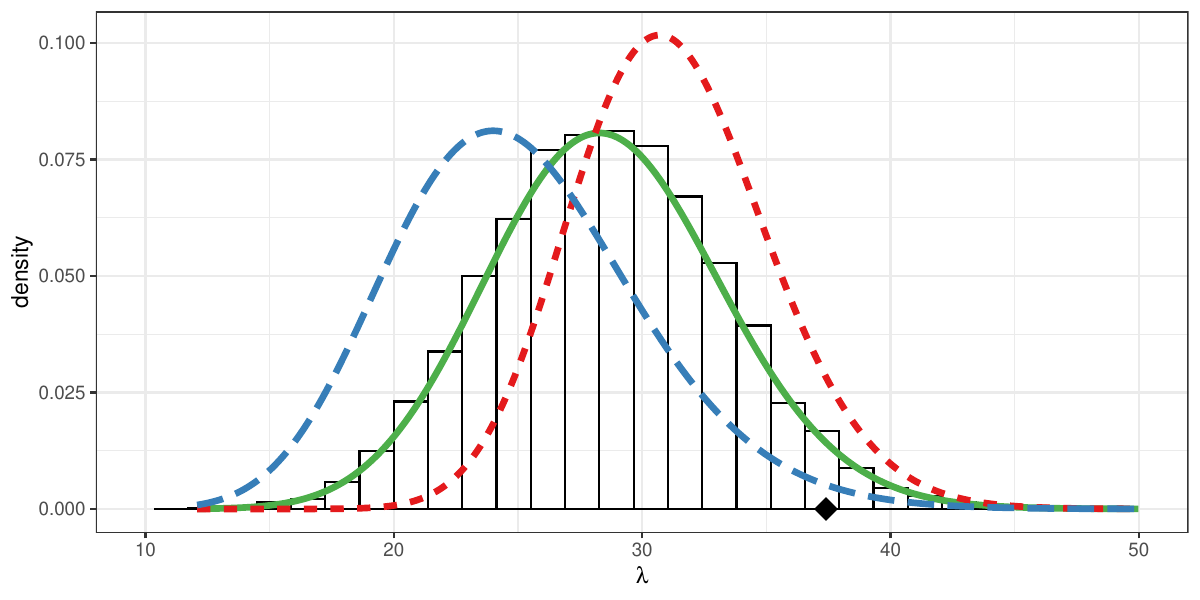}
\caption{Algorithm~\ref{algorithm:abc} produces exact draws (black histogram, $N = 10^4$) from the true posterior (green density), which is different from the na\"ive posterior (red dotted density) which treats the observed $\sss_\diffp = 37.4$ (black diamond) as if without privatization. Blue dashed density is the prior.}
\label{fig:abc}
\end{figure}

Figure~\ref{fig:abc} depicts both the correct and the na\"ive analyses, with hyperparameters $\alpha = 25, \beta = 1$, privacy loss budget $\epsilon = 0.2$, and $\sss_\diffp = 37.4$. The true analytical posterior (green solid density), normalized via numerical integration, coincides with the differentially private \ABC~posterior histogram tabulated from $10^4$ draws from Algorithm~\ref{algorithm:abc}. The correct analysis differs substantially from the incorrect na\"ive  posterior (red dotted density), which treats $\sss_\diffp$ as if it were an observed confidential query. (The latter posterior amounts to the posterior from the standard  Gamma-Poisson conjugate model.) Compared to the correct posterior, the na\"ive posterior succumbs less to the shrinkage effect imposed by the prior. It assigns a heavier weight of evidence to the observed value of $\sss_\diffp$, more so than it deserves. It is furthermore overly concentrated at the mode, underestimating the posterior uncertainty associated with $\theta$. 
 
Appendix~\ref{app:example-count} displays additional experiments that employ Gamma prior distributions with hyperparameters $\alpha = 2, 5, 50, 75$ and $\beta = 1$.
Worth noting is that when the privatized observation $\sss_\diffp$ appears highly unlikely under the chosen prior (or is \emph{in conflict} with it, in other words), the correct posterior heavily discounts the contribution by $\sss_\diffp$. For example, such is the case when $\alpha = 2$ or $5$, as seen in Figure~\ref{fig:abc2} (a) and (b):  the correct posterior is in close alignment with the prior and differs significantly from the na\"ive posterior. As alluded to in Section~\ref{sec:dpabc}, prior-data conflict presents a challenge for \ABC~algorithms in general, because forward sampling tends to explore the area with higher prior predictive concentration. A realized observation far from that area would result in a low acceptance rate. To see this, Table~\ref{tab:acceptance-abc} reports the average acceptance rates and their standard errors over 20 direct repetitions of Algorithm~\ref{algorithm:abc} under various choices of Gamma priors. In comparison with the observed query, these priors range from congruent to conflicting, as can be seen from the varied differences between $\sss_\diffp$ and its prior predictive expectation: $\mathbb{E}_{\theta}(\sss_\diffp) = \int\sss_\diffp\int\eta_\diffp\left(\sss_{\diffp} \mid \sss\right)\int\pi\left(\sss\mid\theta\right)\pi_0 \left(\theta\right)d\theta d\sss d\sss_{\diffp}$.

\begin{table}[t]
\centering
\caption{Acceptance rate of Algorithm~\ref{algorithm:abc} under various priors\label{tab:acceptance-abc} ; $\sss_\diffp = 37.4$}
\begin{tabular}{rrrr}
  \toprule
  prior: $\theta \sim Gamma(\alpha, 1)$ & prior predictive: $\mathbb{E}_{\theta}(\sss_\diffp)$ & acceptance rate (\%) & s.e. (\%) \\ 
  \midrule
$\alpha = 2$ & 2& 0.09 & 0.02 \\ 
$\alpha = 5$ & 5& 0.21 & 0.06 \\ 
 (Figure~\ref{fig:abc}) $\alpha = 25$ & 25 & 16.24 & 0.35 \\ 
$\alpha = 50$ & 50& 19.83 & 0.31 \\ 
$\alpha = 75$ & 75& 0.64 & 0.07 \\ 
   \bottomrule
\end{tabular}
\end{table}

Maximum likelihood estimation for $\theta$ is carried out as follows. The confidential data likelihood is the Poisson density. Importance sampling as described in Algorithm~\ref{algorithm:mcem} is used to construct estimates for (\ref{eq:ets}) at every iteration of the E-step, followed by an analytical M-step.  Appendix~\ref{app:example-count} describes details of the implementation using three stages of successively more stringent tolerance levels. With $\theta^{(1)} = 1$, the algorithm converges to the maximizer $\hat{\theta} = 37.237$, with observed Fisher information estimated to be $1.582\times 10^{-2}$. If $\sss_\diffp$ were erroneously treated as the confidential data, the MLE for $\theta$ would've been $37.4$, and the observed Fisher information would've been $2.674 \times 10^{-2}$, or $69\%$ larger than the correct estimate, again displaying an underestimation of inferential uncertainty. The reduction of Fisher information content reflects a loss of statistical efficiency induced by the privatization mechanism, and is expected in typical inference problems whenever confidential data are replaced with their privatized counterparts. Details of the above calculations can be found in Appendix~\ref{app:example-count}.

\subsection{Lalonde dataset}\label{sec:lalonde}

The Lalonde dataset \citep{lalonde1986evaluating}  was curated from the the randomized trial of the National Supported Work (NSW) Demonstration and nonexperimental comparison data, for the purpose of studying the efficacy of the job training program on recipients' future earnings. The dataset, with a total of 185 treated and 260 control units, is well-studied in the causal inference and econometrics literatures using regression and propensity matching methods, see e.g. \cite{heckman1989choosing,dehejia1999causal,dehejia2002propensity}. We employ the example here to illustrate a Bayesian analysis that compares the 1978 earnings of the treatment and control groups, if $\epsilon$-differentially private versions of the key descriptive statistics were released instead.

Let $z_{i}$ be the observed indicator for whether subject $i$ received treatment ($z_i = 1$) or control ($z_i = 0$), and $y_{i}$ their earning in 1978 (in $\$1k$).  The full parameter of the model is $\theta=\left(\tau,\mu,\sigma_{t}^{2},\sigma_{c}^{2}\right)$, in which $\tau$ is the difference in average earnings between the treatment and control groups, and is the primary parameter of interest.  We posit independent priors for elements of $\theta$, as well as the sampling model
\[
y_{i}\mid z_{i},\theta\sim N\left(\tau z_{i}+\mu,\sigma_{t}^{2}z_{i}+\sigma_{c}^{2}\left(1-z_{i}\right)\right).
\]
For simplicity's sake, we do not consider additional covariates that distinguish the treatment and control subjects.

\begin{figure}
\centering

\includegraphics[width = \textwidth]{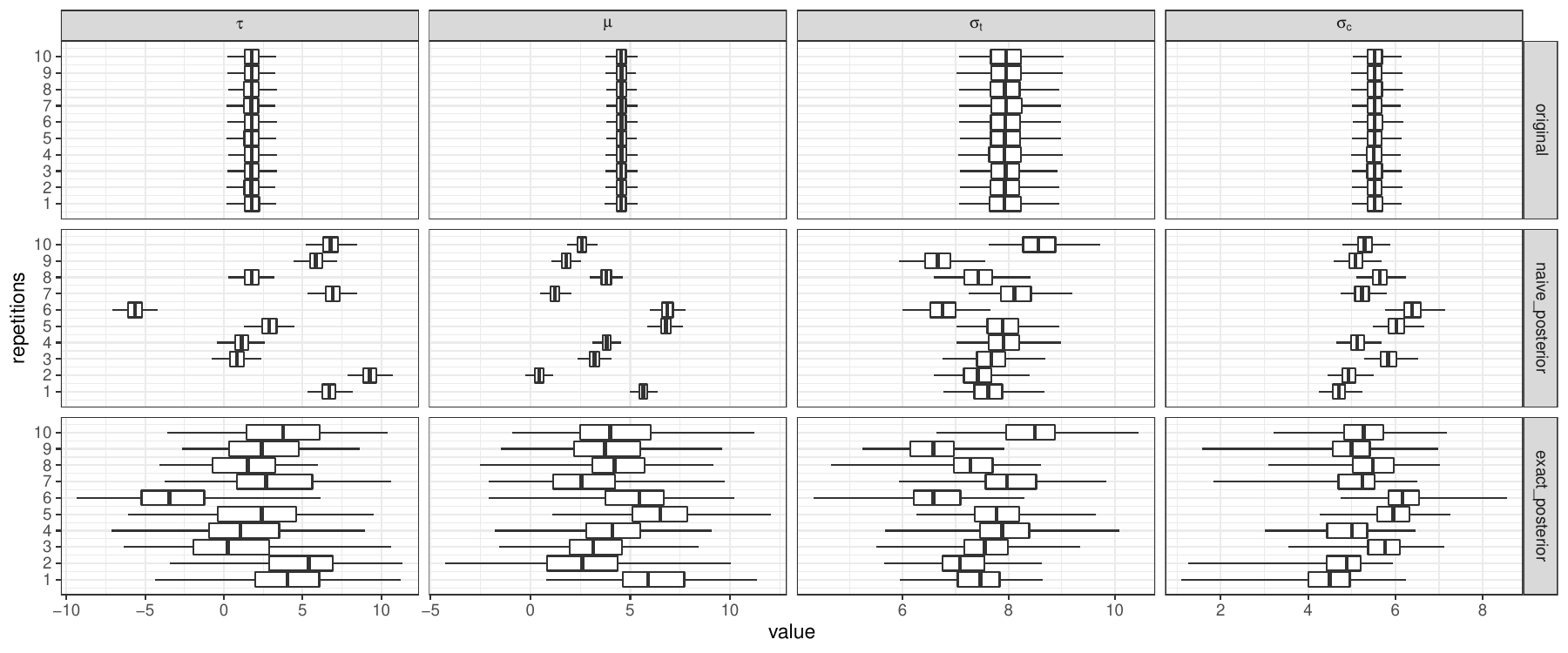}	
\caption{\label{fig:lalonde} Boxplots of $\left(1\%, 25\%, 50\%, 75\%, 99\%\right)$ posterior quantiles of $\left(\tau,\mu,\sigma_{t}^{2},\sigma_{c}^{2}\right)$. Top row: ten repeated {\tt RStan} fittings of the original model to the original data $\sss$; Mid row: na\"ive {\tt RStan} fittings of the original model to ten realizations of $\sss_\diffp$ via the Laplace mechanism; Bottom row: exact posterior fittings using rejection~\ABC~(Algorithm~\ref{algorithm:abc}) on the same ten $\sss_\diffp$ realizations as above.}
\end{figure}

Among  the descriptive statistics that the publisher plans to release,  relevant to the inferential task at hand are the within-group sample means and sample variances: $\sss = \left(\bar{y}_{t}, \bar{y}_{c}, s_{t}^{2},s_{c}^{2}\right)$. Together they make up the sufficient statistic for the full parameter $\theta$.  The top row of Figure~\ref{fig:lalonde} displays the posterior inference for $\theta$ by repeatedly fitting this model in {\tt RStan} using the actual value of $\sss$. Discrepancies among the ten boxplots within each figure, all of them minor, are due to Monte Carlo errors. According to the model, there is a discernible positive treatment effect since the posterior mass of $\tau$ is overwhelmingly positive.

Suppose that the data publisher releases $\epsilon$-differentially private version of sample means and variances. Since the mean and the variance are real-valued functions, they do not have a finite global sensitivity $\Delta_{GS}$ as defined in~\eqref{eq:global}, hence the Laplace mechanism cannot apply directly to them. To circumvent this issue, the publisher may \emph{clamp} the underlying query, that is to enforce its value to stay within a bounded range. For simplicity's sake, suppose that the clamping range on individual income is conservatively set, say to between zero and $\$100k$, and the treatment and control groups are guaranteed to exceed 100 people. This effectively restrict the global sensitivity of $\bar{y}_{t}$ and $\bar{y}_{c}$  to 1 and that of $s_{t}^{2}$ and $s_{c}^{2}$ to 100. For reference, the maximum observed individual income in the dataset is $\$60.3k$, and the treatment and control group are respectively of sizes $n_t = 185$ and $n_c = 260$, ensuring that all confidential query values fall well within the clamping range.  The benefit of conservative clamping is that the privatized statistics would not require truncation correction, even though it amounts to an inefficient privacy budget allocation strategy.   Further suppose two separate privacy loss budgets of $\epsilon = 1/3$ and $100/6$  are respectively expended on the sample means and variances, through Laplace mechanisms employing independent zero-mean noise components with bandwidths $h^{-1} = 1/3$ for each of the sample means  $\bar{y}_{t}, \bar{y}_{c}$, and $h^{-1} = 1/6$ for each of the sample variances $s_{t}^{2}, s_{c}^{2}$.

The middle and bottom rows of Figure~\ref{fig:lalonde} respectively display posterior inferences from na\"ively fitting the original model (i.e. disregarding the privacy mechanism)  in {\tt RStan}, and correctly fitting the exact posterior (i.e. accounting for the privacy mechanism) using rejection~\ABC~of Algorithm~\ref{algorithm:abc}. Both methods were fitted to the same ten independent realizations of $\sss_{\diffp}$ from the Laplace mechanism. Discrepancies among the ten boxplots within each figure in these two rows are due to the random privacy mechanism and to Monte Carlo errors -- the latter to a much lesser extent. We see that with the correct analysis, posterior uncertainty for all parameters are substantially inflated, in part due to the highly inefficient allocation of the privacy loss budget. As a result, we can no longer conclude that the treatment effect is significantly in either direction. However, the posterior quantiles overlap substantially with their counterparts from the original posterior on the top row, indicating that the cost of privacy manifests more as an estimation precision loss rather than bias. This stand in contrast against the na\"ive analysis which delivers tight, yet idiosyncratically displaced, posterior masses. Details of this analysis can be found in Appendix~\ref{app:example-lalonde}.

\section{Conclusion and Discussion}\label{sec:discussion}

Modern likelihood and Bayesian inferences face the challenge of model complexity. They appeal to Monte Carlo and approximate methods to carry out needed computation, even if the resulting inferences are only approximate with respect to the full statistical model. This paper discussed how approximate computation algorithms, specifically~\ABC~and Monte Carlo EM, can be adapted to obtain exact Bayesian and likelihood statistical inferences based on differentially private data products. In both cases, the tuning elements of the approximate computation algorithms are chosen to accord to the specifications of the differentially private perturbation mechanism, which can be made transparent to the data analyst. Both methods are applicable to a wide range of modeling scenarios, and may help data users transition existing methodologies to apply to differentially private data products while maintaining statistical validity of their analysis.

When no privacy mechanism is involved, \ABC~algorithms exhibits a bias whenever it cannot enforce an exact match between the observed and the simulated data \citep{nunes2010optimal,drovandi2011approximate,gleim2013approximate,barnes2012considerate,bernton2019approximate}, which is typically the case in practice. The justification of \ABC~relies on that in the limit as the bandwidth $h \to 0$, the \ABC~posterior $\pi_{\ABC}(\theta \mid \sss)$ approaches the true posterior $\pi(\theta \mid \sss)$ \citep{blum2013comparative,sisson2018handbook}. However in practice,  $h$ cannot be too small in order for the algorithm to generate an adequate number of samples, trading off a larger approximation error with a smaller Monte Carlo error. 

The statistical insight underscored by this paper is the duality
\[
\text{\emph{approximate computation on exact data}} \quad \leftrightarrow \quad \text{\emph{exact computation on approximate data}}.
\]
Differentially private data is approximate data. The perturbation mechanism with which the data were treated serves coincidentally as the attributable cause of the approximation error. When differentially private data are employed, the Monte Carlo error becomes the sole kind of error attributable to the \ABC~algorithm, and vanishes as $N \to +\infty$ as any other consistent method of simulation.  %

The pursuit of differential privacy pits a direct tradeoff against statistical efficiency \citep{duchi2018minimax}. But the efficiency-privacy tradeoff as a statistical consideration is interweaved with the approximation-exactness tradeoff as a computational consideration, a sentiment that is shared by explorations of other simulation-based Bayesian computational algorithms with differentially private data, including stochastic gradient Monte Carlo \citep{wang2015privacy} and Gibbs sampling \citep{foulds2016theory}. For \ABC~algorithms, to insist on maximal statistical efficiency necessitates computational approximation. Whereas the act of data perturbation not only gains differential privacy, but also computational exactness for free. Both the \ABC~and Monte Carlo EM approaches  adapt to differentially private data using the same logic, by setting the tuning parameters governing their numerical performance based on the privacy parameters. Tailoring an algorithm according to the data generative specification exploits the alignment between the statistical and computational tradeoffs, hitting two birds with one stone, so to speak.

There are several computational challenges to the practical implementation of the proposed frameworks. These challenges are of two types: those intrinsic to \ABC~and other forward sampling techniques, and those induced by the privacy mechanism. A weakness in either of these aspects may impact the computational efficiency of these proposals, or in the worst case, render them infeasible. We discuss the two types of challenges below.

A data analyst operating under the dissemination mode of data access is on the receiving end of data products which are designed and privatized by the data curator. As this paper discusses the migration of existing statistical methodology to accommodate privacy-protected data products, we assume that the analyst knows how to perform their preferred analysis on the data product \emph{were it not} privatized, i.e. if the curator releases $\sss$ rather than $\sss_\diffp$. That is, $\pi(\theta\mid\sss)$ in \eqref{eq:noiseless-posterior} is taken to be the ultimate posterior the analyst targets. 
Depending on the model, however, the analyst may or may not prefer to use \ABC~or other forward sampling techniques to draw inference from $\pi(\theta\mid\sss)$. The strength of \ABC~lies in its ability to handle intractable likelihoods, but it presents several limitations. In the construction of the current paper, the intended query function $\sss$ (and hence the private query $\sss_\diffp$) may be multi-dimensional,  where each dimension is generated in isolation, in conjunction, or sequentially. In particular, we do not preclude the identity function,  $\sss(\xx) = \xx$, in which case the privacy perturbation is performed element-wise on the full dataset for publication, such as may be encountered in the \emph{local} differential privacy setting. 

Whenever the full data likelihood does not admit a low-dimensional sufficient summary to $\sss$, the computational efficiency of both proposed algorithms will likely suffer. For classic \ABC, the synthetic data matching step (step 3 in Algorithm~\ref{algorithm:abc}) will be computationally wasteful. The \ABC~literature explores the use of approximate summary statistics \citep{beaumont2002approximate,joyce2008approximately,wegmann2009efficient} to achieve dimension reduction and efficient matching. Unless carefully designed, however, general approximate summary reduction to $\sss$ will complicate the expression of the privacy kernel $\eta_\diffp$, and will destroy the ``exact'' nature of the proposed algorithm. The lack of sufficient reduction challenges the feasibility of other modes of computational for privacy-aware Bayesian inference as well; see e.g. \cite{bernstein2018differentially,bernstein2019differentially}.  The question remains with the data curator: in anticipation of a broad range of data analysis needs, how to choose the query $\sss$ that provides better statistical utility \emph{and} computational efficiency?

Another limitation of \ABC~methods is that their performance depends on the prior and the nature of the state space. As mentioned previously, \ABC~must work with proper prior distributions. This  minimal requirement speaks nothing about the algorithm's efficiency.
As the numerical experiment in Section~\ref{sec:example-count} demonstrates, the acceptance probability of Algorithm~\ref{algorithm:abc} is low when the observed data is in conflict with the prior. The remedy is to devote more sampling resource to areas of the parameter space for which the data exhibit more support. This is a tautology of sorts, since the area we seek is precisely the area with high posterior density, which may be particularly difficult to locate when the parameter space is high dimensional, and when the prior distribution is diffuse (despite being proper).

There are also computational challenges brought forth by the privacy mechanism. Since both proposed algorithms require the transparency of $\eta_\diffp(\sss_\diffp \mid \cdot)$, any act that deprives the analyst's ability to evaluate this quantity also hinders the proposed computational schemes. Two notable causes to diminished transparency of the privacy mechanism are clamping and \emph{post-processing}. As discussed in Section~\ref{sec:lalonde}, the curator performs clamping when the query has unbounded global sensitivity. While na\"ive and conservative clamping (such as presented in Section~\ref{sec:lalonde}) requires little additional work from the analyst, carefully designed clamping procedures typically involve the underlying confidential dataset in a nontrivial fashion \cite[see e.g.][]{biswas2020coinpress}. The resulting privacy mechanism may not be simply captured by an analytically tractable $\eta_\diffp$.  In addition, the {post-processing} of differentially private data products may also complicate an otherwise simple expression of $\eta_\diffp$. Such is the case if the post-processing operation depends nontrivially on aspects of the observed data. For example, the TopDown algorithm imposes \emph{invariants} on the differentially private noisy measurements via optimization-based post-processing \citep{abowd2022topdown}. As a result, the output of the algorithm does not permit a straightforward probabilistic description, which threatens its \emph{congeniality} as a building block in the data processing pipeline \citep{gong2020congenial}. From the statistical point of view, a transparent  privacy mechanism is instrumental to the feasibility of conducting exact statistical inference from privacy-protected data \citep{gong2022transparent}. To ensure transparency of the privacy mechanism is yet another challenging task that lies with the data curator.

\section*{Acknowledgment}
The author wishes to thank Xiao-Li Meng for inspiring discussions, as well as John Abowd, Gary King, Zhiqiang Tan, and three anonymous reviewers for helpful comments. The author gratefully acknowledges research support by the National Science Foundation (DMS-1916002).

\bibliography{master}
\bibliographystyle{abbrvnat}

\newpage
\appendix

\section{Examples of additive perturbation DP mechanisms}\label{app:dp-mechanism}

\begin{example}[$\epsilon$-Laplace mechanism; \cite{dwork2006calibrating}]\label{ex:1}
In \eqref{eq:pert}, let $\uu \sim \text{Lap}_{p}(1)$, the $p$-dimensional product of independent and identically distributed standard Laplace variables, and $h = \epsilon^{-1}\Delta_{GS}(\sss)$, where
\begin{equation}\label{eq:global}
\Delta_{GS}(\sss)=\sup_{\xx,\xx'}\left\{ \left\Vert \sss\left(\xx\right)-\sss\left(\xx'\right)\right\Vert: d\left(\xx,\xx'\right)=1\right\},
\end{equation}
is the global sensitivity of $\sss$, with $\Vert \cdot \Vert$ denoting the $\ell_1$ norm. Then, $\SSS$ is $\epsilon$-differentially private.
\end{example}

\begin{example}[$(\epsilon, \delta)$-Laplace mechanism; \cite{nissim2007smooth}]\label{ex:2}
In \eqref{eq:pert}, let $\uu \sim \text{Lap}_{p}(1)$, $h = \epsilon^{-1}\Delta_\xi(\sss, \xx)$, and $\xi=\epsilon\left\{ 4\left(p+\log\left(2/\delta\right)\right)\right\}^{-1}$, where   
\begin{equation}\label{eq:smooth}
\Delta_{\xi}(\sss,\xx)=\sup_{\xx'}\left\{ e^{-\xi d\left(\xx,\xx'\right)}\Delta_{LS}\left(\sss,\xx'\right):\xx'\in\XX\right\}
\end{equation}
is the {$\xi$-smooth sensitivity} ($\xi>0$) of $\sss$ at $\xx$, and
\begin{equation}\label{eq:local}
\Delta_{LS}(\sss,\xx)=\sup_{\xx'}\left\{ \left\Vert \sss\left(\xx\right)-\sss\left(\xx'\right)\right\Vert:d\left(\xx,\xx'\right)=1\right\}
\end{equation}
is the {local sensitivity} of $\sss$ at $\xx$. Then, $\SSS$ is $(\epsilon, \delta)$-differentially private. 
\end{example}

\begin{example}[Gaussian mechanism; \cite{blum2005practical,nissim2007smooth}]\label{ex:3}
In \eqref{eq:pert}, let $\uu \sim  N({\bf 0}, {\bf I}_p)$ the $p$-dimensional standard multivariate Normal variable, $h = \epsilon^{-1}5\sqrt{2\log(2/\delta)}\Delta_\xi(\sss, \xx)$, and $\xi=\epsilon\left\{ 4\left(p+\log\left(2/\delta\right)\right)\right\}^{-1}$. Then, $\SSS$ is $(\epsilon, \delta)$-differentially private.
\end{example}

The above examples invoke three notions of functional sensitivity (\ref{eq:global})-(\ref{eq:local}), generally denoted as $\Delta(\sss)$, to capture the idea that certain choices of $\sss$ may be more revealing of individual information in $\xx$ than others. The global sensitivity measures the extent to which $\sss$ varies between all conceivable pairs of neighboring datasets, whether or not realized in the observed sample. For example, the global sensitivity of the counting query is $1$. On the other hand, the local sensitivity of $\sss$ measures its maximum variability among neighboring datasets to a given observed dataset $\xx$. The smooth sensitivity strikes a balance between the two, by providing an upper bound on the local sensitivity at $\xx$ in such a way that the bound does not vary too quickly as a function of $\xx$. It is crucial that the scale parameter of the additive perturbation mechanism is chosen as a function of both the sensitivity of $\sss$ as well as the privacy budget, that is, $h = h(\epsilon,\delta,\Delta(\sss))$.

\section{Proof of Theorem~\ref{thm:dp-abc}}\label{app:dp-abc}

\begin{proof}
Let $I$ be the indicator of the event that a draw of $\theta$ is accepted. The joint distribution of all quantities produced by the $i$th iteration is $\tilde{\pi}(\theta,\sss, I) = \pi_0(\theta) \pi(\sss \mid \theta) \tilde{\pi}(I \mid \sss)$, where $\tilde{\pi}(I \mid \sss)$ is the Bernoulli mass function with proportion parameter $c\eta_{\diffp}\left(\sss_{\diffp}\mid\sss\right)$. The marginal distribution of an accepted $\theta$ sample is
\begin{equation}\label{eq:abc-posterior}
	\tilde{\pi}\left(\theta \mid I = 1 \right) 
	= \int\frac{\tilde\pi\left(\theta,\sss,I=1\right)}{\tilde\pi\left(I=1\right)}d\sss
	= \frac{\int\pi_{0}\left(\theta\right)\pi\left(\sss\mid\theta\right)c\eta_{\diffp}\left(\sss_{\diffp}\mid\sss\right)d\sss}{\int\int\pi_{0}\left(\theta\right)\pi\left(\sss\mid\theta\right)c\eta_{\diffp}\left(\sss_{\diffp}\mid\sss\right)d\sss d\theta},
\end{equation}
which is equal to $\pi\left(\theta\mid\sss_{\diffp}\right)$ as defined in (\ref{eq:posterior}). From here, one can see that the overall acceptance probability of Algorithm~\ref{algorithm:abc} is 
\[
\tilde{\pi}(I = 1) = \pi\left(\sss_{\diffp}\right)/\max{\eta_\diffp(\cdot)}.
\]
\end{proof}

Note that under the special case of additive perturbation, the proof of Theorem~\ref{thm:dp-abc} parallels Theorem 1 of \cite{wilkinson2013approximate}. However, there is an important conceptual difference. In \cite{wilkinson2013approximate}, the conditioning query is a query that was observed noiselessly, but construed as if subject to additive error. The \ABC-induced posterior of $\theta$ therein, while essentially identical to (\ref{eq:posterior}), is not the true posterior of $\theta$ but that of a ``best model input $\hat{\theta}$'' given $\sss_\diffp$. With $\sss_\diffp$ being a privatized query, no pretense is necessary in treating it as observed with error, since it indeed was.

\section{Effective sample size for Monte Carlo EM}\label{app:mcem-weights}

In reference to Algorithm~\ref{algorithm:mcem}, at the $t$th iteration, the normalized version of the importance
sampling weights is
\[
\tilde{\omega}_{i}=c_{\left(t\right)}\eta_\diffp\left(\sss_{\diffp} \mid \sss_{i}\right) = c_{\left(t\right)}{\omega}_{i}
\]
where $c_{(t)}=1/\pi(\sss_{\diffp}\mid\theta^{(t)})$ is the reciprocal
of the current approximation to the observed likelihood and is free
of $\sss_{i}$. The weighted estimator $\sum_{i=1}^{N}\tilde{\omega}_{i}b\left(\sss_{i}\right)$ is a consistent estimator of (\ref{eq:ets}) because
\begin{eqnarray*}
\mathbb{E}\left(b\left(\sss\right)\mid\sss_{\diffp},\theta^{\left(t\right)}\right) & = & \int b\left(\sss\right)\pi\left(\sss\mid\sss_{\diffp},\theta^{\left(t\right)}\right)d\sss\\
 & = & \int b\left(\sss\right)\frac{\pi\left(\sss\mid\theta^{\left(t\right)}\right)\eta_\diffp\left(\sss_{\diffp} \mid \sss\right)}{\int\pi\left(\sss\mid\theta^{\left(t\right)}\right)\eta_\diffp\left(\sss_{\diffp} \mid \sss\right)d\sss}d\sss\\
 & = & \int\tilde{\omega}\left(\sss\right) b\left(\sss\right)\pi\left(\sss\mid\theta^{\left(t\right)}\right)d\sss.
\end{eqnarray*}
We have that at the expectation of weights for the $t$th iteration is
\begin{eqnarray*}
\mathbb{E}_{\sss\mid\theta^{\left(t\right)}}\left(\tilde{\omega}\right) 
	&=&	\int c_{\left(t\right)}\eta_\diffp\left( \sss_{\diffp}\mid \sss \right)\pi(\sss\mid\theta^{(t)})d\sss \\
	&=&	c_{\left(t\right)}\pi(\sss_{\diffp}\mid\theta^{(t)})=1,
\end{eqnarray*}
where the subscript $\sss\mid\theta^{\left(t\right)}$ signifies the expectation is evaluated with respect to the current approximation to the latent data likelihood, or equivalently, the proposal distribution of the importance sampler. Similarly,
\begin{eqnarray*}
\textsc{var}{}_{\sss\mid\theta^{\left(t\right)}}\left(\tilde{\omega}\right) & = & \mathbb{E}_{\sss\mid\theta^{\left(t\right)}}\left(\tilde{\omega}^{2}\right)-\mathbb{E}_{\sss\mid\theta^{\left(t\right)}}^{2}\left(\tilde{\omega}\right)\\
 & = & c_{\left(t\right)}^{2}\mathbb{E}_{\sss\mid\theta^{\left(t\right)}}(\eta^{2}_\diffp\left(\sss_{\diffp} \mid \sss\right))-1.
\end{eqnarray*}
This gives rise to the effective sample size
\begin{eqnarray*}
\textsc{ess}^{\left(t\right)}\left(N\right) & = & N/\left(1+\textsc{var}_{\sss\mid\theta^{\left(t\right)}}\left(\tilde{\omega}\right)\right)\\
&=& N\pi^{2}\left(\sss_{\diffp}\mid\theta^{(t)}\right)\mathbb{E}^{-1}_{\sss\mid\theta^{\left(t\right)}}\left(\eta^{2}_\diffp\left(\sss_{\diffp} \mid \sss\right)\right).	
\end{eqnarray*}
See also section 2.5.3 of \citep{liu2008monte}.

\section{Observed score and Fisher information for Monte Carlo EM}\label{app:mcem-details}

We have that the observed data log likelihood
\[
\log\pi\left(\sss_{\diffp}\mid\theta\right)=\log\int\pi\left(\sss_{\diffp}\mid\sss\right)\pi\left(\sss\mid\theta\right)d\sss,
 \]
thus the observed score
\begin{eqnarray*}
\nabla_{\theta}\log\pi\left(\sss_{\diffp}\mid\theta\right) & = & \frac{\int\pi\left(\sss_{\diffp}\mid\sss\right)\nabla_{\theta}\pi\left(\sss\mid\theta\right)d\sss}{\int\pi\left(\sss_{\diffp}\mid\sss\right)\pi\left(\sss\mid\theta\right)d\sss}\\
 & = &\frac{\int\frac{\pi\left(\sss_{\diffp}\mid\sss\right)\nabla_{\theta}\pi\left(\sss\mid\theta\right)}{\pi\left(\sss_{\diffp}\mid\sss\right)\pi\left(\sss\mid\theta\right)}\pi\left(\sss_{\diffp}\mid\sss\right)\pi\left(\sss\mid\theta\right)d\sss}{\int\pi\left(\sss_{\diffp}\mid\sss\right)\pi\left(\sss\mid\theta\right)d\sss}\\
 & = & \int\frac{\nabla_{\theta}\pi\left(\sss\mid\theta\right)}{\pi\left(\sss\mid\theta\right)}\frac{\pi\left(\sss_{\diffp},\sss\mid\theta\right)}{\int\pi\left(\sss_{\diffp},\sss\mid\theta\right)d\sss}d\sss\\
 & = & \int\nabla_{\theta}\log\pi\left(\sss\mid\theta\right)\pi\left(\sss\mid\sss_{\diffp},\theta\right)d\sss \\
 & = & \mathbb{E}\left( \nabla_{\theta}\log\pi\left(\sss\mid\theta\right) \mid\sss_{\diffp},\theta\right).
\end{eqnarray*}
Writing $\lambda_\theta(\sss) = \nabla_\theta\log\pi\left(\sss \mid\theta\right)$, we have that $\mathbb{E}\left( \lambda_\theta(\sss) \mid\sss_{\diffp},\theta^{\left(t\right)}\right)$ serves as the $t$th approximation to the observed score $\nabla_\theta \log\pi\left(\sss_{\diffp}\mid\theta^{(t)} \right)$, giving rise to the expression
\begin{equation*}
 \mathbb{E}\left( \lambda_\theta(\sss) \mid\sss_{\diffp},\theta^{\left(t\right)}\right)\approx 
m\sum_{i=1}^{N}\omega_{i}\lambda_\theta(\sss_i).
\end{equation*}

Similarly the Hessian, or the negative of the observed Fisher information matrix, is
\begin{eqnarray*}	
\nabla_{\theta}^{2}\log\pi\left(\sss_{\diffp}\mid\theta\right)	
	&=&	\int\frac{\nabla_{\theta}^{2}\pi\left(\sss\mid\theta\right)}{\pi\left(\sss\mid\theta\right)}\pi\left(\sss\mid\sss_{\diffp},\theta^{\left(t\right)}\right)d\sss  \\ 
	& &	- \left(\nabla_{\theta}\log\pi\left(\sss_{\diffp}\mid\theta\right)\right)\left(\nabla_{\theta}\log\pi\left(\sss_{\diffp}\mid\theta\right)\right)^{\top} \\
	&=& \mathbb{E}\left(\nabla_{\theta}^{2}\log\pi\left(\sss\mid\theta\right)+\nabla_{\theta}\log\pi\left(\sss\mid\theta\right)\nabla_{\theta}\log\pi\left(\sss\mid\theta\right)^{\top}\mid\sss_{\diffp},\theta\right) \\
	& &	- \left(\nabla_{\theta}\log\pi\left(\sss_{\diffp}\mid\theta\right)\right)\left(\nabla_{\theta}\log\pi\left(\sss_{\diffp}\mid\theta\right)\right)^{\top}.
\end{eqnarray*}
Substituting again $\lambda_\theta(\sss)$ and the expression for the observed score into the above equation, we have that the $t$th approximation to the observed Fisher information $-\nabla_\theta^{2}\log\pi\left(\sss_{\diffp}\mid\theta^{(t)}\right)$  takes the form
\begin{eqnarray*}
& & \mathbb{E}\left(-\nabla_{\theta}\lambda_{\theta}\left(\sss\right)-\lambda_{\theta}\left(\sss\right)\lambda_{\theta}\left(\sss\right)^{\top}\mid\sss_{\diffp},\theta^{\left(t\right)}\right)+\mathbb{E}\left(\lambda_{\theta}\left(\sss\right)\mid\sss_{\diffp},\theta^{\left(t\right)}\right)\mathbb{E}\left(\lambda_{\theta}\left(\sss\right)\mid\sss_{\diffp},\theta^{\left(t\right)}\right)^{\top} 	\nonumber \\
& & \approx  m\sum_{i=1}^{N}\omega_{i}\left\{ -\nabla_{\theta}\lambda_{\theta}\left(\sss_{i}\right)-\lambda_{\theta}\left(\sss_{i}\right)\lambda_{\theta}\left(\sss_{i}\right)^{\top}\right\} + m^{2}\sum_{i=1}^{N}\sum_{j=1}^{N}\omega_{i}\omega_{j}\lambda_{\theta}\left(\sss_{i}\right)\lambda_{\theta}\left(\sss_{j}\right)^{\top} \label{eq:fisher}.
\end{eqnarray*}
See also the appendix of \cite{louis1982finding}.

\section{Details of Section~\ref{sec:example-count}: privatized count inference}\label{app:example-count}

For the Bayesian analysis, by the $\epsilon$-Laplace perturbation mechanism, the conditional distribution of 
$\sss_{\diffp}$ given $\sss$ is $Lap\left(\sss,\epsilon^{-1}\right)$, which
has density $\frac{\epsilon}{2}\exp\left(-\epsilon\left|\sss_{\diffp}-\sss\right|\right)$. By construction, $\sss_\diffp$ is not an integer with probability one, hence
\begin{eqnarray*}
\pi\left(\sss_{\diffp}\mid\theta\right)  &=& \int\pi\left(\sss\mid\theta\right)\pi\left(\sss_{\diffp}\mid\sss\right)d\sss \\ 
    &\propto &  e^{-\theta}\left\{ \sum_{s=0}^{\left\lfloor \sss_{\diffp}\right\rfloor }\frac{\theta^{\sss}}{\sss!}e^{-\epsilon\sss_{\diffp}+\epsilon\sss}+\sum_{s=\left\lceil \sss_{\diffp}\right\rceil }^{\infty}\frac{\theta^{\sss}}{\sss!}e^{\epsilon\sss_{\diffp}-\epsilon\sss}\right\} .
\end{eqnarray*}
Adopting the notations $\theta_{\epsilon}^{+}=\theta e^{\epsilon}$
and $\theta_{\epsilon}^{-}=\theta e^{-\epsilon}$, the first sum within the brackets can be written as
\[
e^{-\epsilon\sss_{\diffp}}\sum_{s=0}^{\left\lfloor \sss_{\diffp}\right\rfloor }\frac{\left(\theta_{\epsilon}^{+}\right)^{\sss}}{\sss!}=e^{\theta_{\epsilon}^{+}-\epsilon\sss_{\diffp}}F_{\theta_{\epsilon}^{+}}\left(\left\lfloor \sss_{\diffp}\right\rfloor \right)
\]
where $F_{\lambda}\left(a\right)$ stands for the $Pois\left(\lambda\right)$
CDF evaluated at $a$. Similarly, the second sum can be written as
\[
e^{\epsilon\sss_{\diffp}}\sum_{s=\left\lceil \sss_{\diffp}\right\rceil }^{\infty}\frac{\left(\theta_{\epsilon}^{-}\right)^{\sss}}{\sss!}=e^{\theta_{\epsilon}^{-}+\epsilon\sss_{\diffp}}\left(1-F_{\theta_{\epsilon}^{-}}\left(\left\lfloor \sss_{\diffp}\right\rfloor \right)\right).
\]
Combining the above with the Gamma prior, $\pi_{0}\left(\theta\right)\propto\theta^{\alpha-1}e^{-\beta\theta}$,
we have that the posterior $\pi\left(\theta\mid\sss_{\diffp}\right)$ takes the form
\begin{equation*}\label{eq:dp-abc-posterior}
\pi\left(\theta\mid\sss_{\diffp}\right)\propto\theta^{\alpha-1}e^{-\left(\beta+1\right)\theta}\left[\frac{\Gamma\left(\left\lceil \sss_{\diffp}\right\rceil ,\theta_{\epsilon}^{+}\right)}{\Gamma\left(\left\lceil \sss_{\diffp}\right\rceil \right)}e^{\theta_{\epsilon}^{+}-{\epsilon\sss_{\diffp}}}+\frac{\gamma\left(\left\lceil \sss_{\diffp}\right\rceil ,\theta_{\epsilon}^{-}\right)}{\Gamma\left(\left\lceil \sss_{\diffp}\right\rceil \right)}e^{\theta_{\epsilon}^{-}+{\epsilon\sss_{\diffp}}}\right],
\end{equation*}
where $\theta_{\epsilon}^{+}=\theta e^{\epsilon}$,
$\theta_{\epsilon}^{-}=\theta e^{-\epsilon}$, $\left\lceil \cdot\right\rceil$ is the ceiling function, and $\Gamma\left(s,x\right)=\int_{x}^{\infty}r^{s-1}e^{-r}dr$ is the incomplete Gamma function with $\Gamma\left(s\right)=\Gamma\left(s,0\right)$ and $\gamma\left(s,x\right) = \Gamma\left(s\right) - \Gamma\left(s,x\right)$.

\begin{figure}[t]
\centering
\qquad (a) $\pi_0 \sim Gamma(2,1)$ \hspace{1.25in} (b) $\pi_0 \sim Gamma(5,1)$
\includegraphics[width = .49\textwidth]{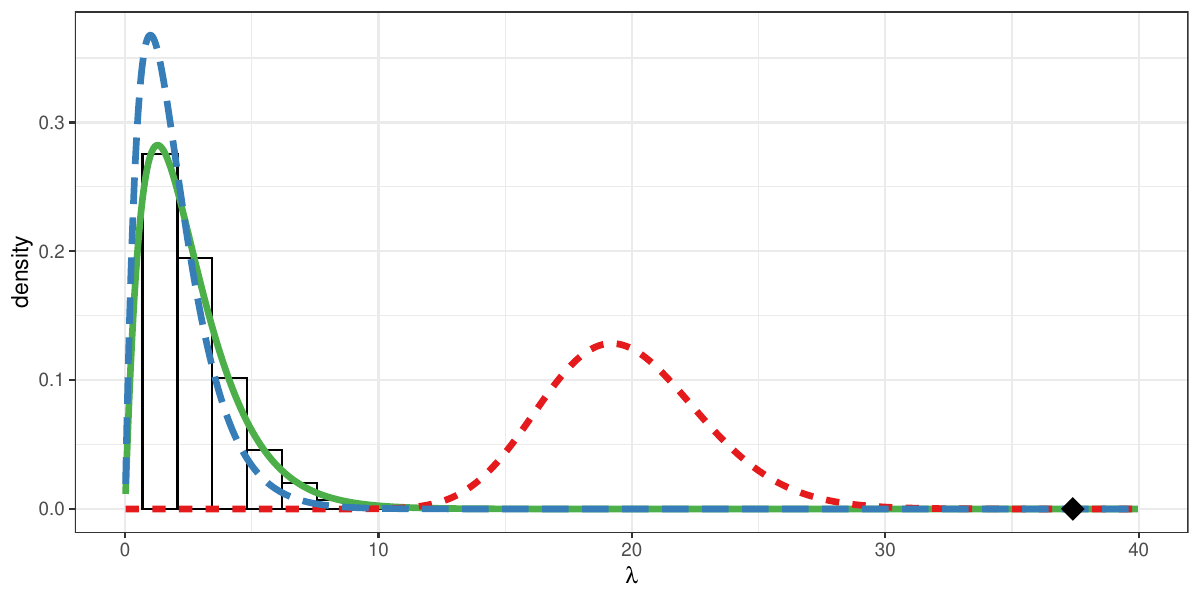}
\includegraphics[width = .49\textwidth]{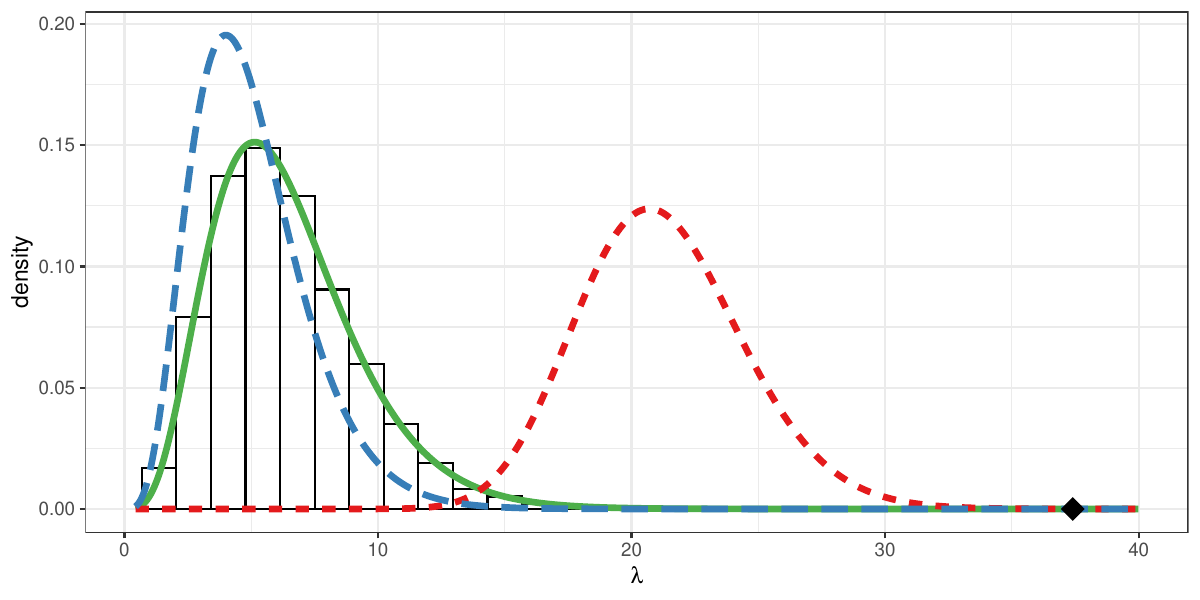}

\qquad (c) $\pi_0 \sim Gamma(50,1)$ \hspace{1.2in} (d) $\pi_0 \sim Gamma(75,1)$
\includegraphics[width = .49\textwidth]{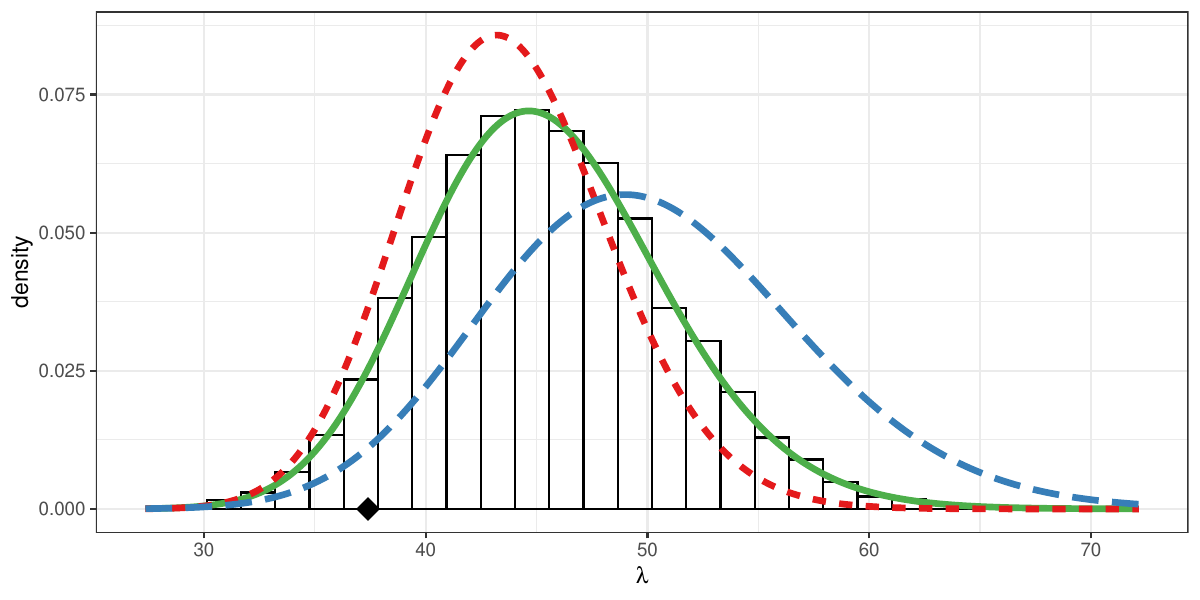}
\includegraphics[width = .49\textwidth]{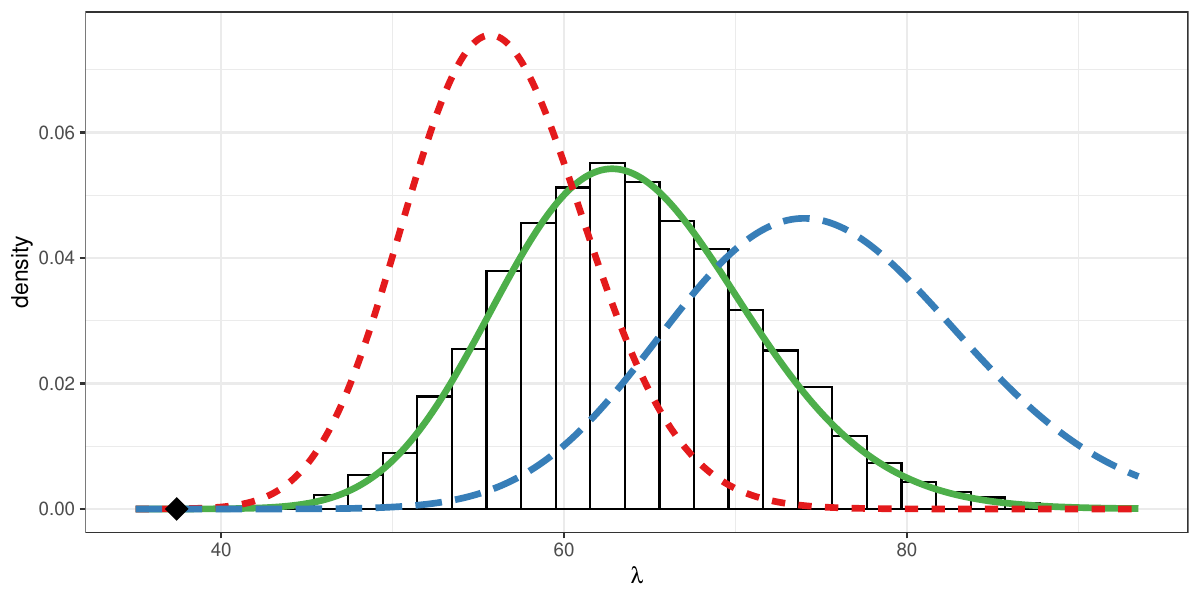}

\caption{Comparisons between the true posterior (green density; approximated by exact draws as black histogram, $N = 10^4$) and the na\"ive posterior (red dotted density) treating observed $\sss_\diffp = 37.4$ (black diamond) as if without privatization, under four different choices of prior distribution $\pi_0$ (blue dashed density).}
\label{fig:abc2}
\end{figure}

Figure~\ref{fig:abc2} displays additional comparisons between the true posterior and the na\"ive posterior for the same privatized count  under other choices of prior distributions in the Gamma family. Notice that when the observed count appears highly unlikely under a chosen prior (such as  $Gamma(2,1)$ or $Gamma(5,1)$), a situation known as \emph{prior-data conflict} \citep{evans2006checking}, the correct posterior heavily discounts the contribution by the privatized observation. The discounting can be seen from the close alignment between the correct posterior (represented by either the solid green density or the black histogram) and the prior (blue dashed density), which in contrast differ drastically from the na\"ive posterior (red dotted density) in Figure~\ref{fig:abc2} (a) and (b). The acceptance rate of Algorithm~\ref{algorithm:abc} in these situations are reported in Table~\ref{tab:acceptance-abc}.

For the implementation of the Monte Carlo EM, three stages of iterations were performed with successively more stringent tolerance levels ($\left|\theta^{\left(t\right)}-\theta^{\left(t-1\right)}\right|< 10^{-3}$, $10^{-4}$, and $10^{-5}$) and larger Monte Carlo sample size ($N = 10^3, 10^5$, and $10^7$), using the stable maximizer from the last stage as the starting point. This is a crude rule to let $N$ increase, hence the Monte Carlo error decrease, as $\theta^{(t)}$ approaches the true \textsc{mle}. Advanced adaptive techniques, such as the ascent-based modification of \cite{caffo2005ascent}, can be employed achieve better performance.

\section{Details of Section~\ref{sec:lalonde}: Lalonde dataset}\label{app:example-lalonde}

Let $z_{i}$ be the indicator variable for whether subject $i$ received
treatment ($z_i = 1$) or control ($z_i = 0$), and $y_{i}$ the earning in 1978 (in
$\$1k$). The full parameter of the model is $\theta=\left(\tau,\mu,\sigma_{t}^{2},\sigma_{c}^{2}\right)$, for which we posit independent priors
\[
\theta\sim\pi_{0}\left(\tau\right)\times\pi_{0}\left(\mu\right)\times\pi_{0}\left(\sigma_{t}^{2}\right)\times\pi_{0}\left(\sigma_{c}^{2}\right),
\]
where for concreteness, we use $\pi_{0}\left(\tau\right) \sim N(0, 5)$, $\pi_{0}\left(\mu\right) \sim N(4, 5)$, $\pi_{0}\left(\sigma_{t}\right) \sim Gamma(2, 0.2)$ and $\pi_{0}\left(\sigma_{c}\right) \sim Gamma(2, 0.2)$ for the analysis. The sampling model is
\[
y_{i}\mid z_{i},\theta\sim N\left(\tau z_{i}+\mu,\sigma_{t}^{2}z_{i}+\sigma_{c}^{2}\left(1-z_{i}\right)\right),
\]
where $\tau$ is the difference in average earnings between the treatment and control groups. Equivalently stated, treatment group earnings have the distribution $N\left(\mu+\tau,\sigma_{t}^{2}\right)$ and the control group earnings have distribution $N\left(\mu,\sigma_{c}^{2}\right)$.

The sufficient statistics for $\theta$ are the within-group mean and sample variances
\[
\sss=\left(\bar{y}_{t},\bar{y}_{c},s_{t}^{2},s_{c}^{2}\right) =\left(\frac{1}{n_{t}}\sum_{i:z_{i}=1}y_{i},\frac{1}{n_{c}}\sum_{i:z_{i}=0}y_{i},\frac{1}{n_{t}-1}\sum_{i:z_{i}=1}\left(y_{i}-\bar{y}_{t}\right)^{2},\frac{1}{n_{c}-1}\sum_{i:z_{i}=0}\left(y_{i}-\bar{y}_{c}\right)^{2}\right).
\]
Due to statistical independence of the sample mean and variance of normal random variables, the likelihood can be equivalently represented by the generative model
\[
\bar{y}_{t},\bar{y}_{c},s_{t}^{2},s_{c}^{2}\mid z,\theta\sim N\left(\mu+\tau,\frac{\sigma_{t}^{2}}{n_{t}}\right)\times N\left(\mu,\frac{\sigma_{c}^{2}}{n_{c}}\right)\times\frac{\sigma_{t}^{2}}{n_{t-1}}\chi_{n_{t}-1}^{2}\times\frac{\sigma_{c}^{2}}{n_{c-1}}\chi_{n_{c}-1}^{2}.
\]

Through a conservative clamping treatment described in Section~\ref{sec:lalonde}, the $\epsilon$-differentially private statistic  $\sss_{\diffp}$ is obtained via a Laplace mechanism with independent Laplace noise components with bandwidth $h^{-1} = (1/3, 1/3, 1/6, 1/6)$ corresponding to $\sss$. Since the clamping range well exceeds the anticipated range of observable data, we do not perform inferential correction for truncation. Both the original analysis using $\sss$ (top row of Figure~\ref{fig:lalonde}) and the na\"ive analysis using $\sss_\diffp$ (bottom row of Figure~\ref{fig:lalonde}) are carried out in {\tt RStan}, whereas the correct analysis  (middle row of Figure~\ref{fig:lalonde}) is carried out using rejection~\ABC~of Algorithm~\ref{algorithm:abc}.

\end{document}